\documentclass[11pt]{article}
\usepackage[utf8]{inputenc}
\usepackage{textcomp}
\DeclareUnicodeCharacter{B0}{\textdegree}
\usepackage[letterpaper, margin=1.3in]{geometry}
\usepackage[edtable]{lineno}
\usepackage[round]{natbib}
\usepackage{amsmath}
\usepackage{graphicx}
\usepackage{color}
\usepackage[noblocks]{authblk}
\usepackage{setspace}

\usepackage{appendix}

\newcommand{\mb}[1]{\mathbf{#1}}
\newcommand{\mr}[1]{\mathrm{#1}}
\newcommand{\ith}[1]{$i^{\mr{th}}$}
\newcommand{\bs}[1]{\boldsymbol{#1}}

\newcommand{\iidsim}{\stackrel{\mathrm{iid}}{\sim}}

\begin{document}
\bibliographystyle{abbrvnat}

\title{Changing measurements or changing movements? Sampling scale and movement model identifiability across generations of biologging technology}

\author[1,4*]{Leah R. Johnson}
\author[2,3,4]{Philipp H. Boersch-Supan}
\author[5]{Richard A. Phillips}
\author[2,3]{Sadie J. Ryan}
\affil[1]{Department of Statistics, Virginia Tech, Blacksburg, VA 24061, USA}
\affil[2]{Department of Geography, University of Florida, Gainesville, FL 32601, USA}
\affil[3]{Emerging Pathogens Institute, University of Florida, Gainesville, FL 32610, USA}
\affil[4]{Department of Integrative Biology, University of South Florida, Tampa, FL, USA}
\affil[5]{British Antarctic Survey, Natural Environment Research Council, Cambridge, CB3 0ET, UK}

\affil[*]{lrjohn@vt.edu}

\maketitle

\doublespacing

\section*{Summary}
\begin{enumerate}
\item Animal movement patterns contribute to our understanding of variation in breeding success and survival of individuals, and the implications for population dynamics.
\item Over time, sensor technology for measuring movement patterns has improved.
Although older technologies may be rendered obsolete, the existing data are still valuable, especially if new and old data can be compared to test whether a behavior has changed over time.
\item We used simulated data to assess the ability to quantify and correctly identify patterns of seabird flight lengths under observational regimes used in successive generations of wet/dry logging technology.
\item Care must be taken when comparing data collected at differing time-scales, even when using inference procedures that incorporate the observational process, as model selection and parameter estimation may be biased. In practice, comparisons may only be valid when degrading all data to match the lowest resolution in a set.
\item Changes in tracking technology, such as the wet/dry loggers explored here, that lead to aggregation of measurements at different temporal scales make comparisons challenging. We therefore urge ecologists to use synthetic data to assess whether accurate parameter estimation is possible for models comparing disparate data sets before planning experiments and conducting analyses such as responses to environmental changes or the assessment of management actions.
\end{enumerate}

\section{Introduction}
Movement is an integral part of the foraging behavior of many animals \citep{nathan2008movement}, and can account for much of their daily energy expenditure \citep{chai1999maximum}.  Observed patterns of movement are determined not only by evolved behaviors and intrinsic state (age, sex, body condition, etc.), but also by environmental conditions (climate, oceanography, etc), and by prey abundance and distribution \citep{hays2016key}. Understanding how different movement strategies affect foraging success may provide insight into the processes underlying survival and reproductive success and, ultimately, population dynamics. Robust quantification of these patterns and how they change through time is required to meet this goal \citep{crossin2014tracking}.

Over recent decades, advances in animal tracking and biologging technology have provided an enormous amount of increasingly precise measurements of animal movement paths \citep{phillips2007foraging,block2011tracking,hussey2015aquatic,kays2015terrestrial,hays2016key}. Researchers use satellite transmitters or data loggers to collect location information; these provide data at intervals which are often constrained by battery power or memory capacity \citep{fedak2002overcoming,edwards2007revisiting,shillinger2012tagging}. 
In large birds and mammals, particularly in recent years, the deployment of GPS loggers, and, in marine animals, of salt-water immersion or temperature-depth loggers, have generated a wealth of tracking data at high temporal resolution and at relatively low cost \citep{mackley2010free,block2011tracking,scales2016identifying}. However, data collected previously using VHF and satellite transmitters (platform terminal transmitters or PTTs), older GPS and immersion loggers, or by human observers are also available, albeit at coarser spatial and/or temporal resolution \citep[e.g.][]{edwards2007revisiting,froy2015age}.  {The increasing use of tracking and biologging technology has also been accompanied by initiatives to archive, share, and exchange animal tracking data \citep{birdlife2004tracking,kranstauber2011movebank}. This wealth of existing data creates opportunities for informative comparisons between archived and new behavioral data. However, the different recording resolutions add complications both in terms of methodology and interpretation.

With the burgeoning of bio\-logging and other ecological research, detailed observations are now available that span a time period that is relevant to the temporal scales of demographic processes, even for long-lived animals, as well as changes in the Earth's climate \citep{hazen2013predicted,crossin2014tracking}. New research avenues have therefore opened for using biologging data to study how movement patterns may be changing across time, including in response to environmental variation \citep{hays2016key}. Most studies that deploy tracking devices on animals, such as seabirds, are usually aimed at answering broad ecological questions about habitat use and foraging behavior in one or a few successive years, as opposed to describing patterns of movement across time frames longer than a decade \citep[but see][]{bogdanova2014among,carneiro2016consistency}. Consequently, device sampling intervals may be sub-optimal for learning about movement over the longer term in post hoc studies. Assessing whether movement strategies have changed requires robust methods and movement models that allow the synthesis of datasets collected at different temporal scales, with differing accuracy, and often with different research aims at the outset.

There are many ways to describe and quantify movement patterns of animals that depend on the type and quality of available data. Many models of foraging assume that organisms move diffusively, {\em i.e.}, that animals perform uncorrelated
Brownian walks as they search for food 
\citep{johnson1992animal}. However, for most animals the
Brownian assumption is clearly inadequate \citep{turchin}.  Super-diffusive
descriptions of movement, such as L\'{e}vy walks or flights
\citep{shlesinger,viswanathan2010fish,watkins2005towards} or intermittent search strategies
\citep{benichou2006two,benichou2007minimal}, which describe movement as small
jumps interspersed with occasional longer jumps are popular alternatives to standard diffusion models as they allow for more complex patterns.  L\'{e}vy walks, which model movements with step lengths determined by a power-law distribution, were first applied in ecology to describe the foraging strategies of wandering albatrosses, {\it Diomedea exulans} \citep{viswanathan1996levy}, and have since been used
to describe search or foraging strategies across many different biological
systems 
\citep[e.g., see references in][]{edwards2007revisiting}.
They were also shown theoretically to represent optimal search strategies for
revisitable targets when the targets are fractally distributed
\citep{viswanathan1999optimizing}. However, the validity of L\'{e}vy flights as
descriptions of animal movement foraging is hotly debated in the ecological literature 
\citep{buchanan2008mathematical,humphries2010environmental,reynolds2012olfactory,travis2007do,edwards2007revisiting,auger2011sampling,viswanathan2011physics}. For instance, although the initial study on albatrosses indicated a L\'{e}vy pattern of foraging \citep{viswanathan1996levy}, after correcting and augmenting the original data, and utilizing improved statistical methods, a later study by \citet{edwards2007revisiting} showed that the L\'{e}vy flight model was {\it not} supported; instead, flight times were more likely to be
gamma distributed. Subsequent 
studies claim new evidence for L\'{e}vy like behavior in certain
marine predators \citep{sims2008scaling,humphries2010environmental,sims2012levy,hays2012high,focardi2014levy,reynolds2016levy}, and that humans exhibit more complex behaviors \citep{gonzalez2008understanding}. Further studies on albatrosses have concluded that foraging patterns of some (although far from all) individuals are well described by modified L\'{e}vy flights or Brownian movement in various contexts, and further concluded that birds utilizing this method are able to consume considerably more prey than they need to satisfy their own energy requirements \citep{humphries2012foraging}. Thus the evidence on L\'{e}vy flights in nature is decidedly mixed.

The controversy surrounding the Levy foraging hypothesis has focused both on the theoretical justification of this process model, as well as the statistical procedures used to distinguish L\'{e}vy walks from other random walks \citep{benhamou2007many,auger2011sampling,plank2013levy}. A substantial body of literature has dealt with different fitting approaches and goodness-of-fit measures \citep{plank2008optimal,white2008estimating,auger2011sampling,edwards2012incorrect}, however the question of parameter identifiability or - for practical purposes - estimability has received considerably less attention \citep[but see][]{auger2011sampling,auger2016state}.

In this study, we use foraging data from albatross species collected a decade apart to explore how the changes in logger technology (and hence the scale and mode of sampling), modeled distributions (statistical fitting), and the treatment of both data and distributions, may influence the findings, and our ability to infer and compare behavior over time. Our analyses focus on a particular type of data from loggers which detect and record saltwater immersion, providing information on wet and dry periods (so called immersion loggers)  \citep{edwards2007revisiting,mackley2010free}. Although a geographic location was not available in some of the earlier deployments, PTT or GPS data have been collected concurrently with the immersion data in the past 20 years, providing improved insights into movements and habitat use \citep{mackley2010free, scales2016identifying, carneiro2016consistency}. Previously, a major consideration was memory capacity, which led to alternative ways of sampling and storing data, which were aggregated at different time-scales on the device during the deployment. The aims of our study were to evaluate model and parameter identifiability for different generations of immersion loggers, using synthetic datasets reflecting different sampling regimes. We further investigated parameter estimation for actual data collected in the wild.


\section{Methods}

In order to determine to what extent the inference of underlying foraging patterns is  influenced by the data collection method, we combine a simulation study with an analysis of two suites of data on flights and water landings at sea collected a decade
apart, 1992-1993 and 2002-2004, from wandering {\it Diomedea exulans} and black-browed albatrosses (\emph{Thalassarche melanophris}). Since these data comprise segments of behaviors that have variously been referred to as steps, trips, tracks, flights, etc., a glossary is provided in Table \ref{tb:gloss}. 

\begin{table}
\caption{A glossary of terms describing movement paths used in this study.}\label{tb:gloss}
\begin{tabular}{|l p{0.80\textwidth} |} 
 \hline
\textbf{Term}& \textbf{Definition}\\
\hline
Trip & A trip is assumed to be one foraging excursion, beginning when the animal leaves the nest site and ending when it returns. A trip is comprised of flights interspersed with (water) landings. \\ 
\hline
Flight & Flights are the subcomponents of a trip, the units of space or time between prey capture attempts, in which the bird is actively flying. \\ 
\hline
Step & In tracking studies of terrestrial animals, this is more commonly used to describe distance, rather than time, and again, represents the sub-unit of a trip. Here, we use interchangeably with flight.\\
\hline
Segment & A discrete time unit over which the wet/dry status of the bird is measured. These segments may be aggregated into longer intervals. \\ 
\hline
Interval & The period over which aggregation of one or more wet-dry segments occurs. In the interval, the number of wet and dry segments are recorded.  Flights are comprised of integer numbers of consecutive completely dry intervals. For data at high (time) resolution data the segment and interval time scales may be the same. \\ 
\hline
Record & The counts of flight lengths (in intervals) within or across trips.\\
 \hline
\end{tabular}

\end{table}

\subsection{Inference Procedure}

All flights are assumed to come from one of four possible distributions: (shifted) exponential; (shifted) gamma; (shifted) q-exponential; pareto. Details of the distributions are given in Appendix \ref{ap:dists}. These true flights are then resampled with (real or virtual) data loggers that discretize or aggregate the flights. 

We use two approaches to infer the parameters of the underlying process from the data. One is to take a ``naive'' maximum likelihood approach, i.e., ignore the observational process, and instead assume that the observed data are drawn without noise from the underlying distribution. Another approach is to use a multinomial maximum likelihood approach that explicitly models the observational process \citep{edwards2007revisiting}. 
In this case, the log-likelihood of the parameters $\boldsymbol\theta$, given a record $\mathbf{r}$ (a set of observations, see Table \ref{tb:gloss}) takes the general form

\vspace{-0.75cm}
\begin{equation}
\ell(\bs{\theta}|\mb{r}) = \sum_{j=1}^{J}d_j \log[\mr{P}(j|\bs{\theta})] 
\end{equation}
where $d_j$ is the number of recorded flights of length $j \in [1, J]$, and $\mr{P}(j|\bs{\theta})$ is the probability of observing a flight of length $j$ given the underlying flight time distribution and observation process. 

One assumption of this multinomial model is that there is some biological lower limit to the possible flight in terms of the length of time spent dry. For instance, if a bird extended its foot out of the water to scratch its head, the logger would record that event as a dry interval; however, ideally, these events would be excluded from any analysis of flights. Following \citet{edwards2007revisiting,reynolds2016levy} and others, we use a lower limit of flight duration (part of an overall trip) of 30 seconds, on the biological assumption that this is not likely to be a flight to a different food patch. This lower limit to flight time is built into the exponential, gamma, and q-exponential distributions as a shift, and into the pareto as the lower set point (See Appendix B for details).

\subsection{Simulation Studies}

We used a suite of simulations to explore the effect of the different logger sampling schemes  (specifically the time scales over which data are aggregated) on our ability to correctly infer parameters values of a known model and to choose the true model if we treat it as unknown.

First we generated a series of ``true'' flights drawn directly from the known distributions  without the observation process. For each of the four distributions, we specified four parameter sets for a total of 16 underlying flight time distributions. When possible, we chose parameters so that the theoretical means between the 4 sets of parameters corresponded between distributions, to ensure that the scales of the processes were comparable. For each of the 16 flight distributions we created 10 simulated data sets of length 3000 (i.e., 10 sets of 3000 flights). 

For each of the 160 simulated data sets, we then ``observed'' the data using our two most extreme sampling regimes, that is intervals of either 1 hour or 30 seconds corresponding to the sampling intervals used in field deployments in 1992 and 2004, respectively \citep{edwards2007revisiting}. More details of the algorithm are in Appendix \ref{ap:obs_method}. This resulted in 320 simulated data sets of length $\leq 3000$ (as the aggregation can result in a subset of flights being labeled as non-flights and thus discarded). These simulated data were used in the following two simulation studies. 
 
\subsubsection*{Parameter Identifiability}
Using the 320 simulated data sets, we attempted to infer the parameters from the underlying flight-time model corresponding to the one that generated the data. We used both a naive maximum likelihood estimate (MLE) (i.e., one that excluded the observational process) and the multinomial with the appropriate observational process. For instance, if the underlying model was an exponential, we fit the exponential model, only. The inferred parameters were then compared to the true parameters that generated the flights.

\subsubsection*{Model Identifiability}
Using the 320 simulated data sets, we fit all four of the possible flight distribution models using of the exact (multi-nomial) likelihood. For each of the 320 data sets, we calculated the Akaike Information Criterion \citep[AIC,][]{akaike1973information,bozdogan1987model}, the Bayesian Information Criterion \citep[BIC,][]{raftery1986choosing,schwarz1978estimating}, and the approximate model probabilities based on BIC \citep[][and given by Eqn.~(\ref{eq:BICprobs}), below]{burnham2004multimodel}, and used these to select the best model for each data set.

\subsection{Immersion Data Analysis}
Observational data on flights and water landings were obtained from immersion loggers deployed on the legs of individual wandering albatrosses and black-browed albatrosses from Bird Island, South Georgia (54$^\circ$ 00'S, 38$^\circ$ 03'W). Multiple types of loggers were deployed between 1992 and 2004. The data are summarized in Table \ref{tb:deployments}. 
\begin{table}[ht!]
\centering
\begin{tabular}[t]{|l|c|c|c|c|c|}
\hline
Study  & Species & Year & Agg. interval (s) & $N_{deployments}$ & $N_{flights}$\\
\hline
BBA2002 & \emph{Thalassarche melanophris} & 2002 & 600 & 1 & 1503 \\
\hline
walb2004 & \emph{Diomedea exulans}  & 2004 & 10 & 39 & 3604\\
\hline
walb1998 & \emph{Diomedea exulans}  & 1998 & 15 & 17 & 878 \\
\hline
walb1993 & \emph{Diomedea exulans}  & 1993 & 720 & 11 & 298\\
\hline
walb1992 &\emph{Diomedea exulans}  & 1992 & 3600 & 21 & 340\\
\hline
\end{tabular}
\caption{Overview of immersion logger datasets used in this study \label{tb:deployments}}
\end{table}

\subsubsection*{Flight length calculation from immersion data}

Wet/dry records were parsed at the highest temporal resolution for each type of logging device, before flight lengths were calculated by merging consecutive time periods recorded as dry (Supplementary Materials \ref{ap:parse}).

\subsubsection*{Model fitting and model selection}
All four flight distributions under the the multinomial likelihood with appropriate discretization and aggregation parameters, were fit to all sets of data noted in Table \ref{tb:deployments}.  Models for each data set were ranked via BIC. Further, approximate model probabilities $P(M_i)$ \citep[based on BIC,][]{burnham2004multimodel} were calculated as: 

\vspace{-0.75cm}
\begin{align}
P(M_i) \approx & \frac{e^{-\frac{1}{2}BIC(M_i)} }{\sum_{r=1}^R
  e^{-\frac{1}{2}BIC(M_r)}} \nonumber\\
&=\frac{e^{-\frac{1}{2}[BIC(M_i)-BIC_{min}]} }{\sum_{r=1}^R
  e^{-\frac{1}{2}[BIC(M_r)-BIC_{min}]}} \label{eq:BICprobs}
\end{align}
where $R$ is the total number of models being considered, and $BIC_{min}$ is the minimum BIC value across those models. The second expression is more numerically stable, and so is the one we use in our calculations

\subsubsection*{Comparing the flight length distributions between years and species}
After selecting the best fitting model via BIC, we examine model fit by plotting the theoretical quantiles verses the observed data quantiles. We then compared the estimates of parameters for our three data sets. In the current likelihood framework we obtain point estimates for all parameters. We can then use these parameters to estimate the means/medians and variances between the fitted models.

\section{Results}

\subsection{Simulation Studies}
 
\subsubsection*{Parameter Identifiability}
Using the 320 simulated data sets, we attempted to infer the parameters from the flight-time model corresponding to that which generated the data using both the naive likelihood (excluding observational process) and the exact multinomial likelihood with the appropriate observational process. Results for the exact likelihood are shown in Figure \ref{f:params_exact_lik} and the naive likelihood in \ref{f:params_naive_lik}. 

Overall, parameter estimates were much more precise for the data recorded at high frequency (i.e., 10 sec sampling), regardless of the underlying true distribution or the scale of the true process. This is because the true lengths of dry periods are recorded with high resolution when data are recorded at this high frequency. However, even for the higher resolution data, using the naive likelihood can bias parameter estimates for some cases of the Pareto and q-exponential distributions. This is probably because even the small amount of truncation of dry periods of 30 seconds or less changes the expected ratio of small flights to longer flights, biasing the estimates. Using the exact likelihood helps to account for these shifts. 

In contrast, estimating parameters for a 1 hour integration step is difficult even when using the exact likelihood if the mean/median flight times are on the order (or less) of the integration period (i.e. if true flight times are less than $\sim 5$ hours in our simulations). Again, use of the exact likelihood can provide better results, although for the Pareto and q-exponential both approaches perform poorly if the true median flight duration is short and the integration interval is long. 

\begin{figure}[h!]
\includegraphics[width=\textwidth, trim=60 0 60 0, clip=TRUE]{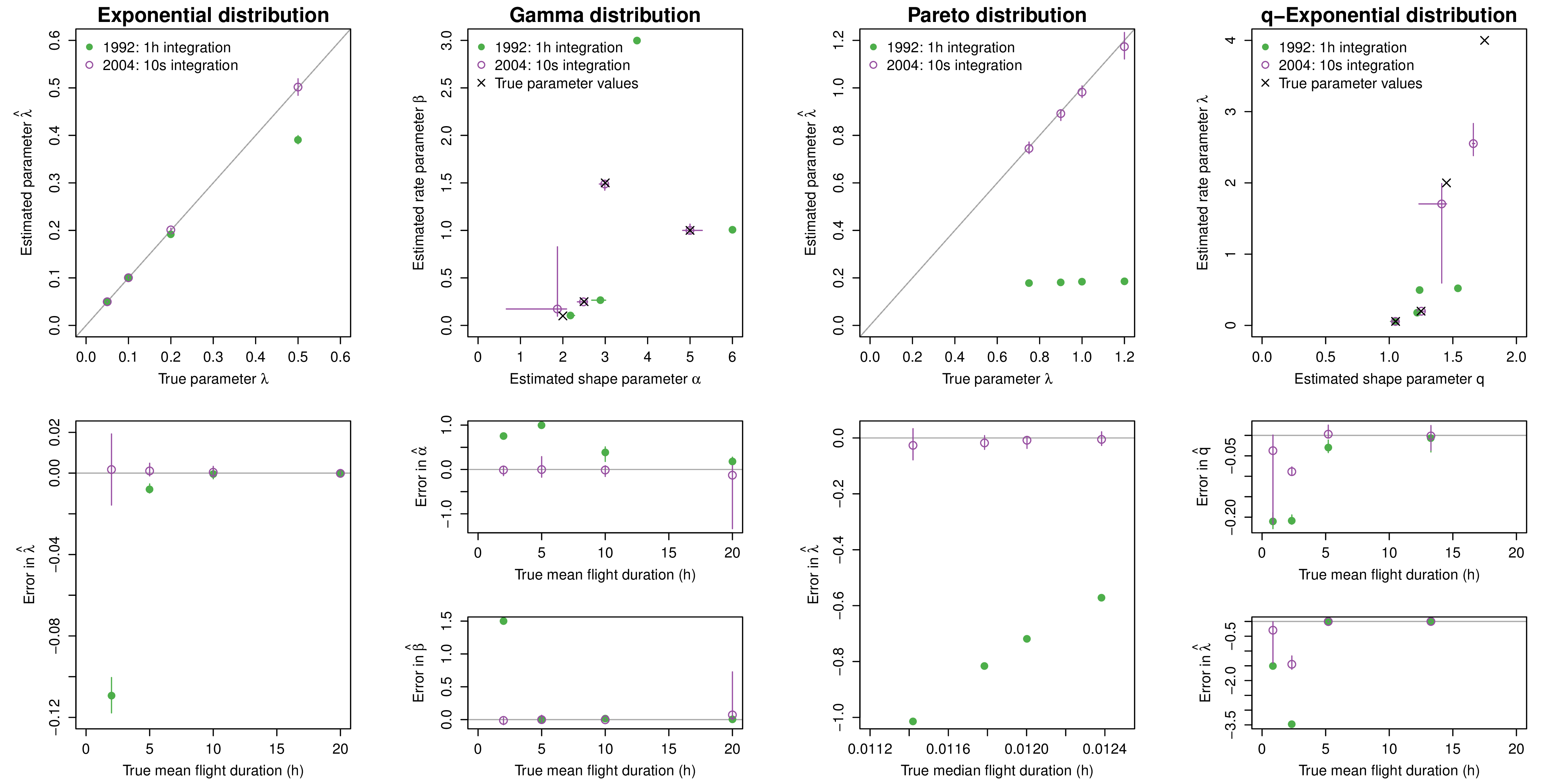}
\caption{Back-estimation of simulated step-length data sets parameters under emulated sampling regimes of two wet/dry activity logger models using \textbf{exact likelihood}. Bars indicate ranges of parameter estimates. \label{f:params_exact_lik}}
\end{figure}

\begin{figure}[h!]
\includegraphics[width=\textwidth, trim=60 0 60 0, clip=TRUE]{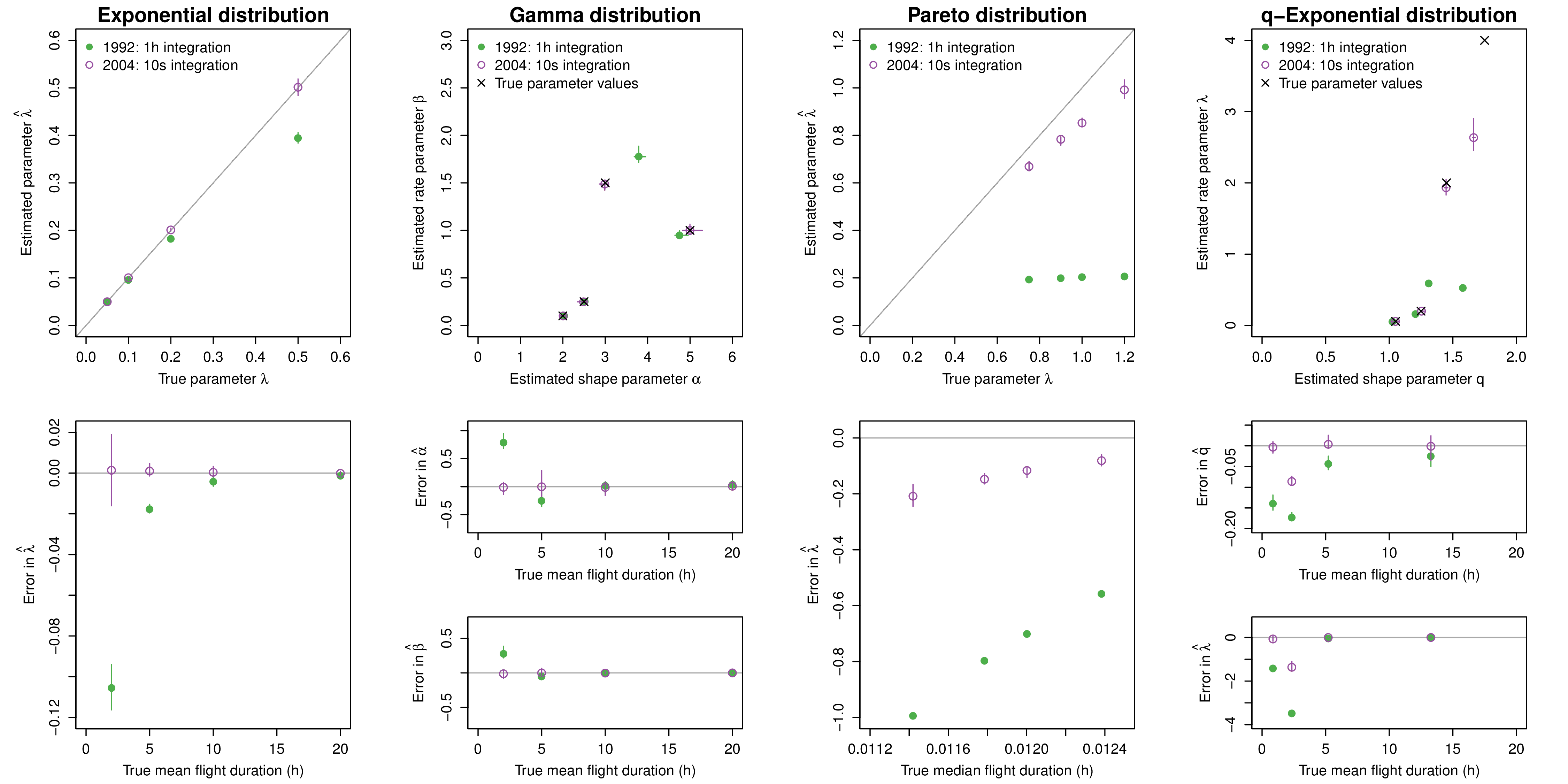}
\caption{Back-estimation of simulated step-length data sets parameters under emulated sampling regimes of two wet/dry activity logger models using \textbf{naive likelihood}. Bars indicate ranges of parameter estimates.}\label{f:params_naive_lik}
\end{figure}

\subsubsection*{Model Identifiability}
Using the 320 simulated data sets, we fit all four of the possible flight distribution models. Focusing on the results from the multinomial approach applied to the appropriate observational process, In Figure \ref{f:sim_mod_ID} we show the proportion of times that the model is identified correctly under the two observation schemes across all 4 flight distribution models. The selected model is indicated when the true model is incorrectly identified. 

\begin{figure}[ht!]
\includegraphics[width=\textwidth,, trim=10 0 0 0, clip=TRUE]{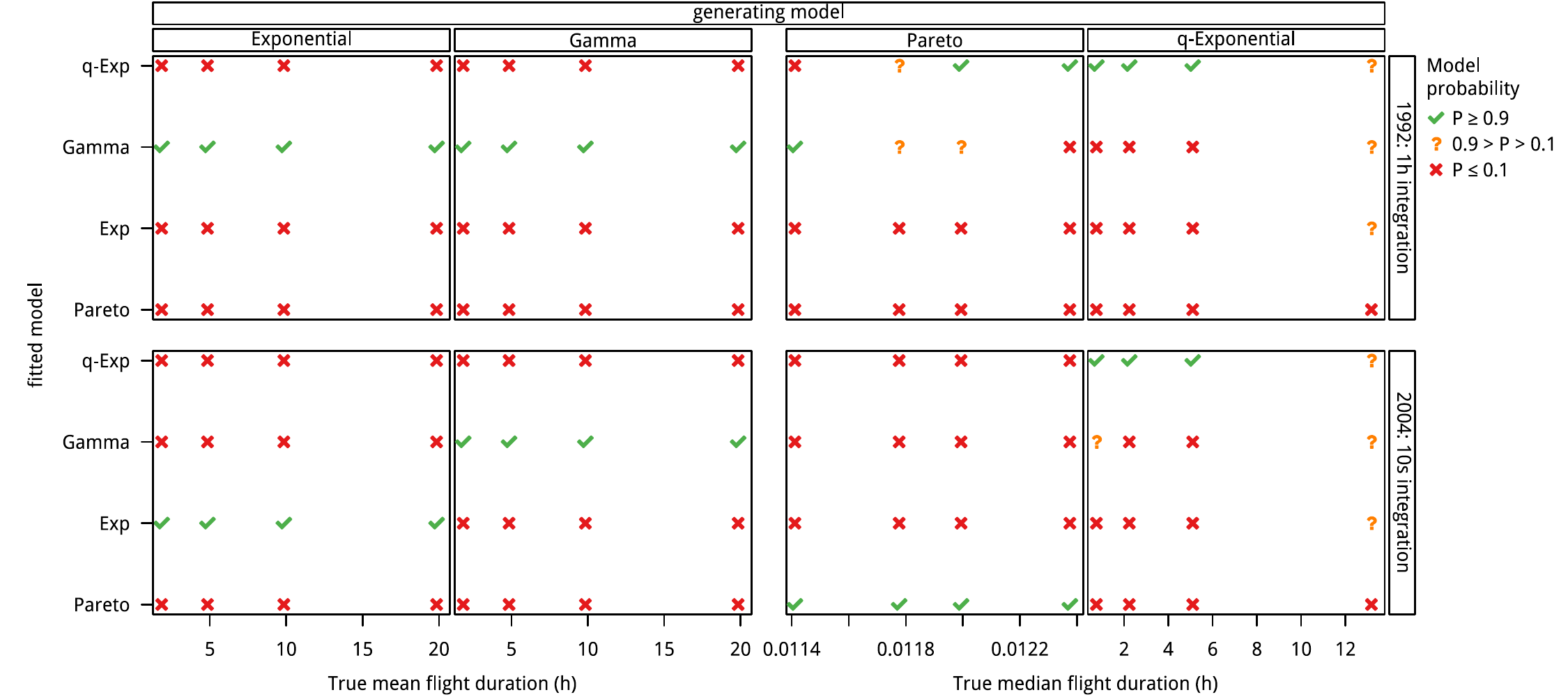}
\caption{\label{fig:model-id} Model identification analysis: Average model probability across the different generating flight time distributions, true mean/median flight lengths, and the low- and high-resolution observation schemes.} \label{f:sim_mod_ID}
\end{figure}

For the short sampling interval (10 sec), the true model almost always has the highest probability. The exception is the q-exponential model, for which there is not a consistent best model if the true median flight length is one hour. Different patterns are apparent for the long (1 hr) integration interval. In this case, only for data generated from the gamma distribution does the true model have high model probability across all parameter settings. For the q-exponential, the true model is chosen if the true median flight time is >1hr (like in the short integration case). If the true underlying models are either Pareto or exponential the true model is never selected. Instead, data from the exponential distribution are always classified as gamma, and the data from the Pareto are either classified as q-exponential or as gamma. These results are congruent with the results from the parameter identifiability simulation experiment: when parameters can be well estimated, that model is likely to be correctly identified as the true model, whereas if the parameter estimates are poor, the patterns in the data that result from the observation process are no longer consistent with the true model, and are better described be one of the alternative models.  

\subsection{Immersion Data Analysis}

\subsubsection*{Flight length calculation from immersion data}

\begin{table}[ht!]
\centering
\begin{tabular}[t]{|l|l|l|l|}
\hline
\textbf{BBA2002}  & $\Delta$BIC & MP\\
\hline
exp & 81.0 & 0\\
\hline
pareto & 3626.9 & 0\\
\hline
gamma & 0 & 1\\
\hline
qexp & 108.0 & 0\\
\hline
\end{tabular}
\hspace{0.5cm}
\centering
\begin{tabular}[t]{|l|l|l|l|}
\hline
\textbf{walb2004}  & $\Delta$BIC & MP\\
\hline
exp & 2689.4 & 0\\
\hline
pareto & 1946.3 & 0\\
\hline
gamma & 0 & 1\\
\hline
qexp & 896.0 & 0\\
\hline
\end{tabular}
\hspace{0.5cm}
\centering
\begin{tabular}[t]{|l|l|l|l|}
\hline
\textbf{walb1998}  & $\Delta$BIC & MP\\
\hline
exp & 1604.6 & 0\\
\hline
pareto & 269.9 & 0\\
\hline
gamma & 0 & 1\\
\hline
qexp & 513.5 & 0\\
\hline
\end{tabular}
\caption{Model selection for observational data.  Here we show the difference in BIC from the best performing model ($\Delta$BIC), such that the best model has a value of 0. We also show the calculated model probabilities, based on Equation (\ref{eq:BICprobs}). Dataset identifiers correspond to Table~\ref{tb:deployments}.\label{tb:obs_bic}}
\end{table}

Across both species, and irrespective of the observation regime, the model that is most consistent with the observed data is the gamma distribution (Table \ref{tb:obs_bic}). This is in line with previous results on a subset of the data \citep{edwards2007revisiting}. Based on gamma Q-Q plots (Figure \ref{fig:model-fit}) for all three of the data sets, the fitted gamma distributions appear to be reasonable for data both from black-browed albatrosses in 2002 and wandering albatrosses in 2004, although both exhibit heavier tails than would be expected from the gamma distribution. The fit for the data from wandering albatrosses in 1998 is much poorer, and is under-estimating the number of short flights. Because the fit is relatively poor, directly assessing whether or not the patterns are consistent across time periods, or comparing between species, should be approached with care. 

\begin{figure}[ht!]
\includegraphics[width=\textwidth]{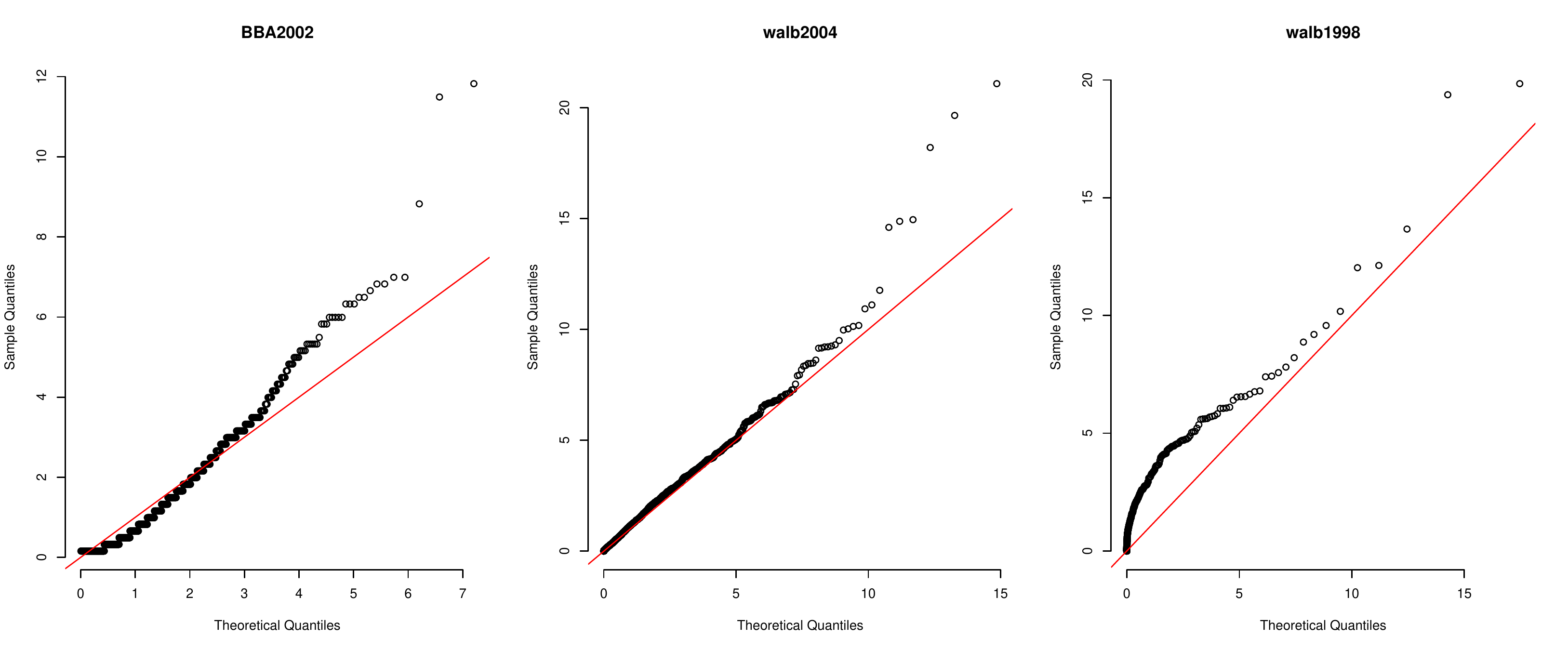}
\caption{\label{fig:model-fit} Gamma QQ plots to assess model fit to the observational data. Dataset identifiers correspond to Table~\ref{tb:deployments}.} 
\end{figure}


\begin{table}[ht!]
\centering
\begin{tabular}[t]{|l|c|c|c|c|}
\hline
\textbf{Gamma}  & $\alpha$ (shape) & $\beta$ (rate) & mean & variance\\
\hline
BBA2002 & 1.38 (1.29 -- 1.47 ) & 1.15 (1.06 -- 1.24) & 1.20 & 1.05 \\
\hline
walb2004 & 0.314 (0.293 -- 0.334) & 0.392 (0.363 -- 0.422) & 0.799 & 2.04\\
\hline
walb1998 & 0.0730 (0.0385 -- 0.107)  & 0.170 (0.133 -- 0.206) & 0.430 & 2.53\\
\hline
\end{tabular}
\caption{Estimated parameters (with approximate 95\% confidence intervals) for the gamma distribution for each of the 3 observational data sets. Additionally, we show the calculated theoretical mean and variance based on the fitted parameters.
Dataset identifiers correspond to Table~\ref{tb:deployments}. \label{tb:gam_fits}}
\end{table}

\section{Discussion}

Quantification of the movements of animals, such as seabirds, provides insights into foraging and migration behavior, the underlying drivers of movement, how movement and behavior may change over time in response to these drivers, and the consequences for individual performance and population dynamics \citep{crossin2014tracking,hays2016key}. The continuing development of new biologging technology for monitoring animal movement has greatly increased the resolution and quality of the data available, increased sample sizes, and reduced the effort required in the field, particularly for obtaining long time-series. These data represent invaluable archives for reconstructing historical movement patterns of animals for comparison with more recent observations. They provide a window into the past for understanding animal movements and the influence of changing environmental conditions, including the abundance and distribution of prey \citep{seco2016distribution,pereira2017using}. As animal movement databases grow \citep{birdlife2004tracking,kranstauber2011movebank}, so do the opportunities for historical comparisons. However, as the resolution and accuracy of tracking devices has changed over time, these comparisons must be made with care to ensure robust interpretation. 

In this paper, we used simulated data to assess the ability to quantify and correctly identify patterns of flight lengths under observational regimes that correspond to the range of older and more recent immersion logger technology for recording landings of foraging seabirds at sea. These simulation experiments are the optimal approach for evaluating new statistical methods -- if it is impossible to reconstruct the true parameters from a known distribution, then the inference method will almost certainly be unreliable when applied to experimental or observational data. Further, this approach allows us to examine the impact of the observational method on the resulting conclusions, and to identify ways of comparing and combining disparate data sets to maximize their value. 

Using simulated data from a set of four underlying flight length distributions that have been hypothesized to describe the flight distributions of seabirds (exponential, gamma, Pareto (corresponding to a L\'{e}vy flight), q-exponential) that are then ``observed'' using a sampling regime typical of immersion loggers deployed in the field, we were able to test the effectiveness of our statistical methods. In particular, we focused on the extent to which incorporating truncation and aggregation, which are part of the sampling procedure, into the likelihood estimation was necessary in order to determine accurate parameter values and correctly identify the underlying model. The results indicated that the inference procedure using the exact likelihood performs as well or better than the naive likelihood in all cases, i.e., accounting for the observation process improves our ability to both identify the model and estimate parameters. This improvement comes with a computational cost, as evaluating the exact likelihood is slower, and some tuning of the maximization procedure is required for individual data sets to achieve convergence. For low and medium resolution data (or if a Pareto distribution is considered, regardless of the resolution) the computational costs are worthwhile. For the very high resolution data (at least every 10 secs.) available from loggers in recent years, it may be sufficient to use the naive likelihood, with no need to incorporate the observational process.

Even when using the likelihood that incorporates the observational process, care must be taken when analyzing data that have been aggregated at time scales that are much longer than the events of interest. In these cases, the parameters can be significantly biased, and a model different from that used to generate the data may be chosen as the best model. In practice, it may therefore be impossible to accurately compare flight patterns from loggers that provide aggregated data at coarse scales from those that provide fine scale data. Instead, the latter may need to be degraded (reaggregated) to the coarser scale to determine if patterns from the two regimes are at least consistent, even if it is not possible to determine if the parameters are the same. This also puts a constraint on the biological questions that may be compared between data taken at different resolutions. For instance, short scale inferences about foraging intervals within a food patch may be unreliable whereas inferences about longer scale movement between patches may be accessable.

Based on the simulated data, we were not able to identify the process model underlying the ``observations'' with the coarse sampling regime consistently and accurately. Thus, we must be cautious when attempting to infer whether or not this particular aspect of the foraging strategy of the albatross has changed over the past two decades based the type of data at hand. Even for the higher temporal resolution data that we present here, the lack of model fit indicated by the QQ plots (Fig.~\ref{fig:model-fit}) is concerning. In particular, there are more long flights than would be typical for the best fitting gamma model. The question is why would this be the case?  In some cases, where concurrent location data are available, we may be able to determine that some longer flights may not represent foraging behaviors, and can be excluded. This was the case for a proportion of the data for which we had concurrent location data. The longer flights could also indicate individual birds that are not exhibiting foraging behavior (for instance attempting to fly out of a storm). In the current analysis we have treated the behavioral state as known, such that the wet status corresponds to feeding/handling attempts and dry to flying foraging. Thus we have not utilized a more complex statistical approach, such as state-space modeling \citep{patterson:2008} that can allow concurrent estimation of behavioral state. If the observed longer flight patterns are a result of a separate non-foraging flying, these methods may be useful for identifying them.

Another possibility is that the mismatch between the data and the models is a symptom of inter-individual differences in behavior, or otherwise more complex behavior than the simple models here allow. If this is the case improving the models themselves, as well as the statistical techniques to analyze them, will be a more fruitful way forward. Even in the case of developing new models for the underlying behavior, it may be that direct parameterization of all model components is not possible. Instead, the quantified patterns explored here could be directly compared with model outputs, for instance emergent flight lengths from am optimal foraging or individual based model. Although model parameterization and validation of more complex models can be challenging, they can allow us to better understand why we see particular patterns and to better predict how behavior may change into the future.   \\

\section*{Acknowledgments}
This project was funded by an NSF grant (PLR-1341649) to L.R.J. and S.J.R. 
We thank Andrew Edwards for sharing his MLE code.

\section*{Author contributions statement}
L.R.J., P.H.B, R.A.P., and S.J.R. conceived the ideas; P.H.B. and S.J.R. processed raw data; L.R.J. and P.H.B. performed data simulations and fitting;  L.R.J., P.H.B, R.A.P, and S.J.R. analyzed the results.  All authors wrote and reviewed the manuscript. 

\section*{Data access} All observational datasets used in this study are available from the Polar Data Centre at the British Antarctic Survey, Cambridge, UK (polardatacentre@bas.ac.uk). Simulated datasets and associated simulation and analysis code will be deposited on http://zenodo.org and assigned a digital object identifier upon acceptance of the manuscript.

\bibliography{movement}

\clearpage
\appendix
\section{Parsing and cleaning of data} \label{ap:parse}
Immersion data (wet/dry records) were downloaded in the field and stored in a variety of file formats. We developed a suite of parsing functions to import these records into R and format the data for further analysis. All code for file parsing and data analysis will be deposited in an online repository.

\subsection{Flight length calculations}
Wet/dry records were parsed at the highest temporal resolution for each type of logging device, before flight lengths were calculated by merging consecutive time periods recorded as dry. In line with previous studies \citep{edwards2007revisiting}, only dry segments with a duration over 30 seconds were counted as flights, to avoid counting preening events and other behaviors that might involve the leg with the logger extended out of the water.

\section{Distributions for step lengths}\label{ap:dists}

We compare 4 distributional models for flight times: shifted exponential, pareto, shifted gamma (which is related to an exponentially truncated pareto), and the shifted q-exponential. Below we give brief details on each of these distributions. Implementations of these distributions in {\sf R} are included in the supplementary materials.

\subsubsection*{Shifted Exponential Distribution}

The shifted exponential is a generalization of the exponential where the support has been shifted some positive amount $x_0$. That is the pdf, $f(x)$ is given by

\vspace{-0.75cm}
\begin{align}
f(x) = 
  \begin{cases}
    0, & \text{for } x < x_0 \\
    \lambda e^{-\lambda (x-x_0)} & \text{for } x \geq x_0
  \end{cases}
\end{align}
where $\lambda>0$ is the rate parameter. 
Similarly was can say that $X-x_0\sim \mathrm{Exp}(\lambda)$.  The theoretical mean of the exponential is $1/\lambda$. If the data were observed directly, the maximum likelihood estimator (MLE) is given by 

\vspace{-0.75cm}
\begin{align}
\hat{\lambda} = \frac{1}{\bar{x}-x_0}
\end{align} 
where $\bar{x}$ is the sample mean.

\subsubsection*{Pareto Distribution}
The Pareto (Type 1) distribution is is power law probability distribution defined above a lower limit, $x_0$. The pdf, $f(x)$ is given by:

\vspace{-0.75cm}
\begin{align}
f(x) = 
  \begin{cases}
    0, & \text{for } x < x_0 \\
    \frac{\alpha x_0^\alpha}{x^{\alpha+1}} & \text{for } x \geq x_0
  \end{cases}
\end{align}
where $x_0>0$ is the minimum possible value (that also determines the scale of the process), and $\alpha>0$ is a unitless shape parameter. This formulation is the same as the  typical power law used in Levy studies, but with the shape parameter redefined so that $\alpha = \mu-1$ so that the parameter is constrained to be positive instead of $>1$.  
The mean and variance of the Pareto are only defined for a subset of parameter values: for $\alpha<1$ the mean approaches infinity, and the variance is also infinite for $\alpha \leq 2$.

In this paper, we assume that we have a biologically defined lower limit, $x_0$, so we are only concerned with estimating the shape parameter, $\alpha$. If the data were observed without error, the MLE for $\alpha$ is given by

\vspace{-0.75cm}
\begin{align}
\hat{\alpha} = \frac{n}{\sum_i(\ln x_i - \ln x_0)} \label{eq:paretoMLE}
\end{align}
where $n$ is the number of data points. If we were also estimating $x_0$, the MLE is simply $\hat{x_0}=\min_i x_i$, and this estimator would be plugged into Eqn.~\ref{eq:paretoMLE}.

\subsubsection*{Shifted Gamma Distribution}
The shifted gamma distribution is a generalization of the gamma where the support has been shifted some positive amount $x_0$. That is the pdf, $f(x)$ is given by

\vspace{-0.75cm}
\begin{align}
f(x) = 
  \begin{cases}
    0, & \text{for } x < x_0 \\
    \frac{\beta^\alpha}{\Gamma(\alpha)} (x-x_0)^{\alpha-1} e^{-\beta (x-x_0)} & \text{for } (x) \geq x_0.
  \end{cases}
\end{align}
where $\alpha>0$ is the shape parameter and $\beta>0$ is the rate parameter.
Similarly was can say that $X-x_0\sim \mathrm{Gamma}(\alpha, \beta)$. The theoretical mean of the standard Gamma distribution in this case is given by $\alpha/\beta$ and the variance by $\alpha/\beta^2$. There are no closed form MLEs for both parameters of the Gamma distribution. Instead we find these numerically.

\subsubsection*{Shifted Q-Exponential Distribution}
The q-exponential distribution is a generalization of the exponential distribution with heavy (possibly power-law) tails. The pdf, $f(x)$ of the shifted distribution is given by

\vspace{-0.75cm}
\begin{align}
f(x) = 
  \begin{cases}
    0 & \text{for } x < x_0 \\
    (2-q)\lambda e_q (\lambda (x-x0)) & \text{for } (x) \geq x_0.
  \end{cases}
\end{align}
where $1\leq q<2$ is the shape parameter, $\lambda>0$ is the rate parameter and 

\vspace{-0.75cm}
\begin{align*}
e_q(x) = [1+(1-q)x]^{\frac{1}{1-q}}.
\end{align*}
When $q=1$ we regain the exponential distribution. Similarly was can say that $X-x_0\sim \mathrm{Pareto}(q, \lambda)$. Here we restrict ourselves to the case where $1\leq q<2$ as this is the range of $q$ for which the support of the distribution is on $[0; \infty)$. As with the gamma, we determine the MLE for the parameters analytically. 

\section{Multinomial likelihoods for the the biologger data}\label{ap:lik}

As in \citet{edwards2007revisiting}, we use a multinomial maximum likelihood approach to estimate the parameters of the underlying flight process while taking into account the observational process of the biologgers. Here we briefly present the likelihoods described in detail in the Supplementary Materials 1 and 2 from  \citet{edwards2007revisiting}, with the equations generalized slightly to account for differences in sampling protocols. 

Most generally, the log-likelihood of the PDF parameters $\boldsymbol\theta$, given a set of observations $\mathbf{r}$ take the general form

\vspace{-0.75cm}
\begin{align}
\ell(\boldsymbol\theta|\mathbf{r}) = \sum_{j=1}^{J}d_j \log[P(j|\boldsymbol\theta)] 
\label{eq:llik}
\end{align}
where $J$ is the number of recorded flights of length. The form of the multinomial probability $P(j|\theta)$ depends on both the underlying continuous distribution as well as the observation protocol. We recognize and implement likelihood functions for two classes of observation protocols: 1) discretized only (2004-type likelihood from \citet{edwards2007revisiting}) and 2) discretized and aggregated (1992-type likelihood from \citet{edwards2007revisiting}). Both forms of the likelihood, as well as functions for creating data, are implemented in {\sf R} and included as part of the supplemental materials. 

\subsection{Discretized only data (2004-type)}

Data of these types consist of sequences of wet/dry indicators within short intervals of length $s$ seconds (10 sec in the wandering albatross data from 2004). The record, $R$, is defined to be the number, $j$, of consecutive dry readings in between two wet readings (see Section \ref{ap:parse} for further details). We define $m$ as the minimum interval, in seconds, that constitutes a flight (e.g., 30 sec for the data from 2004). The minimum record length that we include as a flight in our data set is $m/s+1$, as records shorter than this can include flights that are shorter than $m$. Following \citet{edwards2007revisiting}), we can write the probability of observing flights of length $j$ as:

\vspace{-0.75cm}
\begin{align}
\mathrm{P}(R=j|\theta) =  (1-j) & \int_{s(j-1)}^{sj} f(x; \theta)dx +
(1+j)\int_{sj}^{s(j+1)} f(x; \theta)dx  \nonumber \\
& + \frac{1}{s} \left[\int_{s(j-1)}^{sj} x f(x; \theta)dx +
\int_{sj}^{s(j+1)} x f(x; \theta)dx \right]
\label{eq:prob_j_2004}
\end{align}
where $f(x; \theta)$ is the pdf of the underlying flight time distribution. 

Following \citet{edwards2007revisiting}), we exclude records of length $j_{\mathrm{min}}=m/s$ as these can correspond to flights of less than $m$. However, we must account for the dry intervals of $>m$ that we miss by excluding the records of length $j_{\mathrm{min}}$. The probability of obtaining a record of length $j_{\mathrm{min}}$ is given by

\vspace{-0.75cm}
\begin{align}
\mathrm{P}(R=j_{\mathrm{min}}|\theta) =  
(1+j_{\mathrm{min}})\int_{sj_{\mathrm{min}}}^{s(j_{\mathrm{min}}+1)} f(x; \theta)dx +
 \frac{1}{s}
\int_{sj_{\mathrm{min}}}^{s(j_{\mathrm{min}} + 1)} x f(x; \theta)dx \label{eq:prob_jmin_2004}
\end{align}

We can then obtain the full likelihood for our data by inserting Eqn.~\ref{eq:prob_j_2004} into Eqn~\ref{eq:llik} and subtracting $n$ (the number of records) times the log of 1 minus Eqn.~\ref{eq:prob_jmin_2004}:

\vspace{-0.75cm}
\begin{align}
\ell(\boldsymbol\theta|\mathbf{r}) =  \sum_{j=1}^{J}d_j & \left( (1-j) \int_{s(j-1)}^{sj} f(x; \theta)dx +
(1+j)\int_{sj}^{s(j+1)} f(x; \theta)dx \right. \nonumber \\ 
 &  \left. + \frac{1}{s} \left[\int_{s(j-1)}^{sj} x f(x; \theta)dx +
\int_{sj}^{s(j+1)} x f(x; \theta)dx \right] \right)  \nonumber \\
& - n \log\left( 1-P(R=j_{\mathrm{min}}|\theta) \right).
\end{align}

The pdf, $f(x; \theta)$, from any of the probability distributions described in Appendix \ref{ap:dists} can be used in this log likelihood. This function is then minimized numerically to obtain estimates of the parameters of the distribution. 

\subsection{Discretized and Aggregated data (1992-type)}

Given memory limitations in the older types of immersion loggers, data were aggregated. The loggers deployed on wandering albatrosses in 1992 tested for saltwater immersion every three seconds, and if at least half of the tests were positive in segments of length $\epsilon$ (15 sec for the wandering albatross data from 1992), the segment was counted as wet. The segments of length $\epsilon$ were then aggregated over a larger interval $s$ (1 hour), where only the number of $\epsilon$-length segments that were wet in $s$ are recorded. Thus for each interval of length $s$ a number of segments that are wet will be an integer number on $[0; s/\epsilon]$ ($s/\epsilon = 240$ for the 1992 WALB data). Since it is impossible to discern the exact number or pattern of immersion events within the interval, the exact flight times cannot be distinguished. Thus, the record $R$ for this case is defined to be the number, $j$, of consecutive completely dry intervals in between two intervals with at least one immersion. As before we define $m$ as the minimum interval, in seconds, that constitutes a flight (e.g., 30 sec for the data from 1992). Thus the minimum flight length is shorter than the minimum record length, which is by definition 1. 

Analogous to the previous section, and following \citet{edwards2007revisiting}) we can write the probability of obtaining a record of length $j$:

\vspace{-0.75cm}
\begin{align}
\mathrm{P}(R=j|\theta) = \int_{sj}^{s(j+1)}(x-j)f(x; \theta)dx 
+ \int_{s(j+1)}^{s(j+2)}(2-x+j)f(x; \theta)dx .
\end{align}
Similarly to the in the previous section, this equation is included in the expression for the multinomial likelihood probability and subtracted from the portion relating to $\mathrm{P}(R=0|\theta)$ to get the full log-likelihood:

\vspace{-0.75cm}
\begin{align}
\ell(\boldsymbol\theta|\mathbf{r}) =  \sum_{j=1}^{J}d_j & \left( \int_{sj}^{s(j+1)}(x-j)f(x; \theta)dx 
+ \int_{s(j+1)}^{s(j+2)}(2-x+j)f(x; \theta)dx \right)  \nonumber \\
& - n \log\left( 1-\mathrm{P}(R=0|\theta) \right).
\end{align}
As for the previous section, and distribution of interest can be inserted, and the log likelihood is then maximized to obtain estimates of the parameters for the underlying flight time distributions.

\subsection{Sampling from the multinomial data distribution}\label{ap:obs_method}

Obtaining samples from the multinomial data distribution is straightforward, as for both observation methods (discretized or discretized and aggregated) a flight is the number of intervals that are fully dry.

\begin{enumerate}
\item Take a ``true'' flight time draw from a known distribution, $\tau$.
\item Randomly choose the start time of the flight uniformly within the observation interval. That is, the start time is $ t_{s} \iidsim \mr{U}(0, s)$ where $s$ is the length of the interval, for instance 10 sec for the 2004 WALB data or 1 hour for the 1992 WALB data.
\item The length of the flight, in segments, is then calculated as $\tau_{\mr{obs}}=\mr{floor}((\tau-t_{s})/s)$.
\item If $\tau_{\mr{obs}} > m$ (i.e., of a minimum length to be counted as a flight), that flight is added to the record, $R$.
\end{enumerate}

\end{document}